\documentclass[aps,showpacs,preprintnumbers,amsmath,amssymb,superscriptaddress,floatfix,a4paper,nofootinbib,11pt,floatfix,nofootinbib,onecolumn]{revtex4}
\usepackage{graphicx}
\usepackage{bm}
\usepackage{rotating}
\usepackage{array}
\usepackage{xcolor}
\usepackage{amsmath,amssymb}
\usepackage{mathrsfs}
\usepackage{graphicx}
\usepackage{color}
\usepackage{subfigure}
\usepackage{fancyhdr}
\usepackage{multirow}
\usepackage{float}
\usepackage{epsfig}
\usepackage{amsfonts}
\usepackage{bm}
\usepackage{multirow}
\usepackage{nicefrac}
\newcommand{\ba}{\begin{eqnarray}}
\newcommand{\ea}{\end{eqnarray}}

\usepackage{bbm}
\newcommand{\be}{\begin{eqnarray}}
\newcommand{\ee}{\end{eqnarray}}

\begin{document}

\title{Warm deformed $R^{2}$ inflation}

\author{Apirak Payaka} 
\email{apirak.pa@mail.wu.ac.th}
\affiliation{College of Graduate Studies, Walailak University, Thasala, Nakhon Si Thammarat, 80160, Thailand}
\affiliation{School of Science, Walailak University, Thasala, \\Nakhon Si Thammarat, 80160, Thailand}

\author{Waluka Amaek} 
\email{waluka.am@gmail.com}
\affiliation{College of Graduate Studies, Walailak University, Thasala, Nakhon Si Thammarat, 80160, Thailand}

\author{Phongpichit Channuie} 
\email{channuie@gmail.com}
\affiliation{College of Graduate Studies, Walailak University, Thasala, Nakhon Si Thammarat, 80160, Thailand}
\affiliation{School of Science, Walailak University, Thasala, \\Nakhon Si Thammarat, 80160, Thailand}

\date{\today}

\begin{abstract}

In this work, we study warm inflationary scenario based on a deformation of $R^{2}$ gravity. We start considering $R^{p}$ and assume $p=2(1+\delta)$ with $\delta\ll 1$ so that we simply obtain warm $R^{2}$ inflation when setting $\delta=0$. We then derive the potential in the Einstein frame and consider a dissipation parameter of the form $\Gamma = C_{1}T$ with $C_1$ being a coupling parameter. We focus only on the strong regime of which the interaction between inflaton and radiation fluid has been taken into account. We also consider a detailed analysis of the background dynamics, considering the evolution of the relevant quantities. We compute inflationary observables and constrain the parameters of our model using latest observational data reported by Planck. From our analysis, we discover that with proper choices of parameters the derived $n_s$ and $r$ are in good agreement with the Planck 2018 observational constraints. Particularly, we constrain the potential scale $U_{0}$ of the models.

\end{abstract}


\maketitle


\section{Introduction}
A framework so called cosmic inflation responsible for an early rapid expansion of our Universe becomes a pillar of modern cosmology. It is successful not only to describe important issues that plague the standard Big Bang model, e.g. the horizon and	flatness problems, but also provides a dynamical mechanism for generating the primordial energy density perturbations seeding for a late time large scale structure. This was well known as \lq\lq cold inflation\rq\rq\, \cite{Starobinsky:1980te,Sato:1980yn,Guth:1980zm,Linde:1981mu,Albrecht:1982wi}. In the standard picture, the (p)reheating mechanism at the end of inflation is required in order to have particles/radiation populating the universe. These involve the presence of interactions between the inflaton with other fields resulting the (partial) decay of the inflaton into ordinary matter and radiation, see e.g. \cite{Linde:2005ht,Albrecht:1982mp,Abbott:1982hn}.

However, an alternative approach that the (p)reheating is unnecessary was later proposed. The process can be reliable if one introduces a coupling between inflaton and radiation of which the energy density of radiation can be maintained almost a constant during inflation. The mentioned alternative
scenario was known as \lq\lq warm inflation\rq\rq\,\cite{Berera:1995wh,Berera:1996fm, Berera:1999ws,Berera:2008ar,Bartrum:2013fia}. Such a scenario gained much attention to the community. In other words, it was originally proposed to provide sufficiently hot thermal bath. In the context of warm inflation, it was found that recent studies in many different theories were proposed. For instance, the authors of Ref.\cite{Dymnikova:2000gnk} conducted a possible realization of warm inflation owing to a inflaton field self-interaction. Additionally, models of minimal and non-minimal coupling to gravity were investigated in Refs.\cite{Panotopoulos:2015qwa,Benetti:2016jhf,Motaharfar:2018mni,Graef:2018ulg,Arya:2018sgw,Kamali:2018ylz}. Recently, warm scenarion of the Higgs-Starobinsky (HS) model was conducted \cite{Samart:2021eph}. The model includes a non-minimally coupled scenario with quantum-corrected self-interacting potential in the context of warm inflation \cite{Samart:2021hgt}. An investigation of warm inflationary models in the context of a general scalar-tensor theory of gravity has been made in Ref.\cite{Amake:2021bee}.

In this work, we investigate warm inflationary models in the context of a deformation of $R^{2}$ gravity. We introduce a coupling between inflaton and radiation -- a dissipative term. We demonstrate that the model can complete the radiation dominated Universe at the end of inflation and confront the predictions with the last Planck satellite data.

The paper is organized as follows: In Section \ref{s2}, we will take a short recap of the formalism in the $R^{p}$ theory with $p\geq 2$. Here we present detailed derivations of the field equations as well as the potential in the Einstein frame. All relevant dynamical equations in warm inflation under the slow-roll approximation are given in Section \ref{s3}. In Section \ref{s4}, we consider the deformed $R^{2}$ scenario and derive the spectral index and the tensor-to-scalar ration of the model. In section \ref{s5}, we compare the results in this work with the observational data. Finally, we conclude our findings in the last section.

\section{$R^{p}$ Setup}\label{s2}
One of the simplest classes of a modification to Einstein gravity is to engineer the Einstein-Hilbert term in the action. One possibility is a generic function of the Ricci scalar. This class of theories is well known as the $f(R)$ theories. There were much earlier and pioneer works on $f(R)$ and other gravity theories, see \cite{Nojiri:2010wj,Nojiri:2017ncd}. In this section, we consider the traditionally $4$-dimensional action in $f(R)$ gravity including the matter fields and closely follow setup given in Refs. \cite{Sotiriou:2008rp,DeFelice:2010aj}. 
\ba
S = \frac{1}{2\kappa^2}\int d^4x \sqrt{-g}f(R) + \int d^4x \sqrt{-g}{\cal L}_{M}(g_{\mu\nu},\Psi_{M})\,,
\label{act}
\ea
where we have defined $\kappa^{2}=8\pi G=8\pi/m^{2}_{\rm Pl}=1/M^{2}_{Pl}$, $g$ is the determinant of the metric $g_{\mu\nu}$, and the matter field Lagrangian ${\cal L}_{M}$ depends on $g_{\mu\nu}$ and matter fields $\Psi_{M}$, and $m_{P}$ and $M_{P}$ represent Planck mass and reduced Planck mass, respectively, with $M_{Pl}=m_{Pl}/\sqrt{8\pi}$. The field equation can be directly derived by performing variation of the action (\ref{act}) with respect to $g_{\mu\nu}$ to obtain \cite{Sotiriou:2008rp,DeFelice:2010aj}
\ba
F(R)R_{\mu\nu}(g) - \frac{1}{2}f(R)g_{\mu\nu}-\nabla_{\mu}\nabla_{\nu}F(R)+g_{\mu\nu}\Box F(R) = \kappa^{2}T^{(M)}_{\mu\nu}\,,
\label{eom}
\ea
where $F(R)=\partial f(R)/\partial R$ and the operator $\Box$ is defined by $\Box\equiv (1/\sqrt{-g})\partial_{\mu}(\sqrt{-g}g^{\mu\nu}\partial_{\nu})$. Basically, the energy-momentum tensor of the matter fields is given by a definition $T^{(M)}_{\mu\nu}=(-2/\sqrt{-g})\delta(\sqrt{-g}{\cal L}_{M})/\delta g^{\mu\nu}$. Here it satisfies the continuity equation such that $\nabla^{\mu}T^{(M)}_{\mu\nu}=0$. The action (2.1) in $f(R)$ gravity generally corresponds to a non-linear function $f$ in terms of $R$. It is possible
to derive an action in the Einstein frame under the conformal transformation \cite{Fujii2003,Maeda:1988ab}:
\begin{eqnarray}\label{a1}
{\tilde g}_{\mu\nu}=\Omega^{2}g_{\mu\nu}\,,
\end{eqnarray}
where $\Omega^{2}$ is the conformal factor and a tilde commonly represents quantities in the Einstein frame. The Ricci
scalars $R$ and ${\tilde R}$ in the two frames are related via
\begin{eqnarray}\label{Rt}
R=\Omega^{2}\big({\tilde R}+6{\tilde \Box}\ln\Omega-6{\tilde g}^{\mu\nu}\partial_{\mu}\ln\Omega\partial_{\nu}\ln\Omega\big)\,,\label{Rti}
\end{eqnarray}
where
\begin{eqnarray}
{\tilde \Box}\ln\Omega=\frac{1}{\sqrt{-{\tilde g}}}\partial_{\mu}\big(\sqrt{-{\tilde g}}{\tilde g}^{\mu\nu}\partial_{\nu}\ln\Omega\big)\,,\quad\quad\partial_{\mu}\ln\Omega=\frac{\partial_{\mu}\ln\Omega}{\partial {\tilde x}^{\mu}}\,.
\end{eqnarray}
We rewrite the action (\ref{act}) in the form
\begin{eqnarray}
S^{J}=\int d^{4}x\sqrt{-g}\Big(\frac{1}{2\kappa^{2}}FR-V\Big)+\int d^{4}x\sqrt{-g}L_{M}\big(g_{\mu\nu},\Psi_M\big)\,,\label{Ja}
\end{eqnarray}
where
\begin{eqnarray}
V=\frac{FR-f}{2\kappa^{2}}\,.
\end{eqnarray}
Using Eq.(\ref{Rti}) and the relation $\sqrt{-g}=\Omega^{-4}\sqrt{-{\tilde g}}$, the action (\ref{Ja}) is transformed as
\begin{eqnarray}
S^{E}&=&\int d^{4}x\sqrt{-{\tilde g}}\Big(\frac{1}{2\kappa^{2}}F\Omega^{-2}\big({\tilde R}+6{\tilde \Box}\ln\Omega-6{\tilde g}^{\mu\nu}\partial_{\mu}\ln\Omega\partial_{\nu}\ln\Omega\big)-\Omega^{-4}V\Big)\nonumber\\&+&\int d^{4}x\sqrt{-g}L_{M}\big(g_{\mu\nu},\Psi_M\big)\,.
\end{eqnarray}
The Einstein frame action as a linear action in ${\tilde R}$ can be directly obtained using $\Omega^{2}=F$, and it is very useful to introduce $\kappa\phi=\sqrt{3/2}\ln F$. Then we have $\ln \Omega=\kappa \phi/\sqrt{6}$. Because of the Gauss’s theorem, the integration $\int d^{4}x\sqrt{-{\tilde g}}{\tilde \Box}\ln\Omega$ vanishes. Therefore, the action in the Einstein frame reads
\begin{eqnarray}\label{Ef}
S^{E}&=&\int d^{4}x\sqrt{-{\tilde g}}\Big(\frac{1}{2\kappa^{2}}{\tilde R}-\frac{1}{2}{\tilde g}^{\mu\nu}\partial_{\mu}\phi\partial_{\nu}\phi-U(\phi)\Big)\nonumber\\&+&\int d^{4}x\sqrt{-{\tilde g}}L_{M}\big(F^{-1}(\phi){\tilde g}_{\mu\nu},\Psi_M\big)\,,
\end{eqnarray}
where the scalar degree of freedom takes a
canonical form with a potential
\begin{eqnarray}\label{Ue}
U(\phi)=\frac{V}{F^{2}}=\frac{FR-f}{2\kappa^{2}F^{2}}\,.\label{Po}
\end{eqnarray}
Let us consider inflationary dynamics in the Einstein frame for the scenario also known as the generalized $R^{2}$ model or $R^{p}$ model \cite{Motohashi:2014tra,Liu:2018htf,Renzi:2019ewp,Rojas:2022dky}. With $f(R)=R+\lambda R^{p}$, we find
\begin{eqnarray}\label{phiR}
\phi=\sqrt{\frac{3}{2}}\frac{1}{\kappa}\ln\Big(1+\lambda\,p\, R^{p-1}\Big)\,.
\end{eqnarray}
Substituting the above expression into Eq.(\ref{Po}), we obtain
\begin{eqnarray}\label{Uep}
U(\phi)=\frac{FR-f}{2\kappa^{2}F^{2}}=U_{0}e^{-2\sqrt{2/3}\kappa\phi}\Big(e^{\sqrt{2/3}\kappa\phi}-1\Big)^{\frac{p}{p-1}}\,.
\end{eqnarray}
where
\begin{eqnarray}\label{Uepp}
U_{0}=\frac{1}{2\kappa^{2}}(p-1)p^{p/(1-p)}\lambda^{1/(1-p)}\,.
\end{eqnarray}
Note that for $p=2$ and $\lambda=1/(6M^{2})$ the potential (\ref{Uep}) recovers the potential for $R^2$ inflation:
\begin{eqnarray}\label{Ue}
U(\phi)=\frac{3M^{2}}{4\kappa^{2}}\Big(1-e^{-\sqrt{2/3}\kappa\phi}\Big)^{2}\,.
\end{eqnarray}
Hence the Lagrangian density of the field $\phi$ is given by $L_{\phi}=-\nicefrac{1}{2}{\tilde g}^{\mu\nu}\partial_{\mu}\phi\partial_{\nu}\phi-U(\phi)$ with the energy-momentum tensor
\begin{eqnarray}
{\tilde T}^{(\phi)}_{\mu\nu}=-\frac{2}{\sqrt{-{\tilde g}}}\frac{\delta\big(\sqrt{-{\tilde g}}L_{\phi}\big)}{\delta {\tilde g}^{\mu\nu}}=\partial_{\mu}\phi\partial_{\nu}\phi-{\tilde g}_{\mu\nu}\Big[\frac{1}{2}g^{\mu\nu}\partial_{\alpha}\phi\partial_{\beta}\phi+U(\phi)\Big]\,.
\end{eqnarray}
We notice from Eq.(\ref{Ef}) that the scalar field $\phi$ is directly coupled to matter in the Einstein frame. In order to
see this more explicitly, we take the variation of the action (\ref{Ef}) with respect to the field $\phi$ following the usual Euler-Lagrange technique:
\begin{eqnarray}\label{Ep}
\partial_{\mu}\Big(\frac{\partial\big(\sqrt{-{\tilde g}}L_{\phi}\big)}{\partial_{\mu}\phi}\Big)+\frac{\partial\big(\sqrt{-{\tilde g}}L_{\phi}\big)}{\partial\phi}+\frac{\partial L_{M}}{\partial\phi}=0\,,
\end{eqnarray}
implying that
\begin{eqnarray}
{\tilde \Box}\phi-U_{,\phi} + \frac{1}{\sqrt{-{\tilde g}}}\frac{\partial L_{M}}{\partial\phi} =0\,,\quad{\rm where}\quad {\tilde \Box}\phi=\frac{1}{\sqrt{-{\tilde g}}}\partial_{\mu}\big(\sqrt{-{\tilde g}}{\tilde g}^{\mu\nu}\partial_{\nu}\phi\big)
\end{eqnarray}
The energy-momentum
tensor of matter in the Einstein frame is given by
\begin{eqnarray}
{\tilde T}^{(M)}_{\mu\nu}=-\frac{2}{\sqrt{-{\tilde g}}}\frac{\delta\big(\sqrt{-{\tilde g}}L_{M}\big)}{\delta {\tilde g}^{\mu\nu}}\,.
\end{eqnarray}
Using the standard technique, the derivative of the Lagrangian density $L_{M}=L_{M}(g_{\mu\nu})=L_{M}(F^{-1}(\phi){\tilde g}_{\mu\nu})$ with respect to $\phi$ yields
\begin{eqnarray}
\frac{\partial L_{M}}{\partial\phi}=\frac{\delta L_{M}}{\delta g^{\mu\nu}}\frac{\partial g^{\mu\nu}}{\partial\phi}=\frac{1}{F(\phi)}\frac{\delta L_{M}}{\delta {\tilde g}^{\mu\nu}}\frac{\big(\partial F(\phi){\tilde g}^{\mu\nu}\big)}{\partial\phi}=-\sqrt{-{\tilde g}}\frac{F_{,\phi}}{2F}{\tilde T}^{(M)}_{\mu\nu}{\tilde g}^{\mu\nu}\,.
\end{eqnarray}
In $f(R)$ gravity, we have $-F_{,\phi}/2F=\mathbbm{X}=-1/\sqrt{6}$. It then follows that
\begin{eqnarray}\label{Add}
\frac{\partial L_{M}}{\partial\phi}=\sqrt{-{\tilde g}}\kappa \mathbbm{X}{\tilde T}\,,
\end{eqnarray}
with ${\tilde T}={\tilde g}_{\mu\nu}{\tilde T}^{\mu\nu(M)}=-{\tilde \rho}_{M}+3{\tilde P}_{M}$ in which we have assumed perfect fluids in the Einstein frame. Substituting Eq.(\ref{Add}) into Eq.(\ref{Ep}), we obtain the field
equation in the Einstein frame:
\begin{eqnarray}\label{Add1}
{\tilde \Box}\phi-U_{,\phi} +\kappa \mathbbm{X}{\tilde T}=0\,,
\end{eqnarray}
showing that the field $\phi$ is directly coupled to matter. 

\section{Slow-roll dynamics in warm inflation}\label{s3}
It is worth mentioning that we will directly couple the fermions in the Einstein frame Lagrangian (\ref{Ef}). In the following, we assume the model present in Ref.\cite{Bastero-Gil:2016qru} for the interactions. Considering the Einstein frame action with the flat FLRW line element, the action (\ref{Ef}) leads to the Friedmann equation for warm inflation taking the form
\begin{eqnarray}
H^2 = \frac{1}{3\,M_p^2}\left( \frac{1}{2}\,\dot\phi^2 + U(\phi) + \rho_R\right)\,,
\end{eqnarray}
with $\dot{\phi}=d\phi/dt$ and $\rho_{r}$ being the energy density of the radiation fluid with the equation of state given by $w_{r}=1/3$. As of the standard fashion, the dynamics of the scalar field ($\phi$) with the dissipative term ($\Gamma$) in the context of warm inflation scenario is also governed by the Klein-Gordon equation. It is described via
\begin{eqnarray}
\ddot\phi + 3H\,\dot\phi + U'(\phi) = -\Gamma\,\dot\phi\,,
\end{eqnarray}
where $U'(\phi)=dU(\phi)/d\phi$. The above relation is equivalent to the evolution equation for the inflaton energy density $\rho_\phi$ given by
\begin{eqnarray}
\dot \rho_\phi + 3 H ( \rho_\phi + p_\phi) = - \Gamma ( \rho_\phi +p_\phi) \,,
\label{rhoinf}
\end{eqnarray}
with pressure $p_\phi = \dot \phi^2/2 - U(\phi)$, and $\rho_\phi + p_\phi= \dot \phi^2$. Energy conservation then implies that the
energy lost of the inflaton field must transfer to some other fluid component $\rho_\alpha$. Here the  RHS of Eq. (\ref{rhoinf}) acts as the source term. Hence we have
\begin{eqnarray}
\dot \rho_\alpha+ 3 H ( \rho_\alpha + p_\alpha) = \Gamma ( \rho_\phi +
p_\phi) \,.
\end{eqnarray}
In case of radiation, we have $\rho_\alpha= \rho_R$ and
\begin{eqnarray}
\dot \rho_R + 4 H \rho_R  = \Gamma \dot{\phi}^2\,. \label{eomrad}
\end{eqnarray}
A condition for warm inflation requires $\rho^{1/4}_{R}>H$ in which the dissipation potentially affects both the background inflaton dynamics, and the primordial spectrum of the field fluctuations. To have the accelerated
expansion, the motion of the inflaton field has to be overdamped during warm inflation. Following Refs.\cite{Zhang:2009ge,Bastero-Gil:2011rva}, we consider the general form of the dissipative coefficient, given by
\begin{eqnarray}
\Gamma  = C_{m}\frac{T^{m}}{\phi^{m-1}}\,, \label{Q1}
\end{eqnarray}
where $m$ is an integer and $C_{m}$ is associated to the dissipative microscopic dynamics. Different choices of $m$ have been studied in Refs.\cite{Zhang:2009ge,Bastero-Gil:2011rva,Bastero-Gil:2012akf}. Namely, (1) $m=1$: this case corresponds to the high temperature regime, see Refs.\cite{Berera:2008ar,Panotopoulos:2015qwa,Bastero-Gil:2016qru}; (2) $m=3$: this model is motivated by a supersymmetric scenario \cite{Berera:2008ar,Bastero-Gil:2011rva,Bastero-Gil:2010dgy}, and is found in a minimal warm inflation \cite{Berghaus:2019whh,Laine:2021ego,Motaharfar:2021egj}. Instead of the the Hubble term, this can be achieved due to the present of the extra friction term, $\Gamma$. In slow-roll regime, the equations of motion reduce then to:
\begin{eqnarray}
3 H ( 1 + Q ) \dot \phi &\simeq&  -U_\phi    \,,\label{eominfsl} \\
4 \rho_R  &\simeq& 3 Q\dot \phi^2\,. \label{eomradsl}
\end{eqnarray}
where we have introduced the dissipative ratio $Q=\Gamma/(3 H)$ and $Q$ is not necessarily constant. Concretely, the ratio $Q$ may increase or decrease during inflation since the coefficient $\Gamma$ may depend on $\phi$ and $T$.  The flatness of the potential $U(\phi)$ in warm inflation is measured in terms of the slow roll parameters which are defined in Ref.\cite{Hall:2003zp} given by
\begin{eqnarray}
\varepsilon &=& \frac{M_p^2}{2}\left( \frac{U'}{U}\right)^2\,,\quad \eta = M_p^2\,\frac{U''}{U}\,,\quad \beta = M_p^2\left( \frac{U'\,\Gamma'}{U\,\Gamma}\right)\,.
\label{SR-parameters}
\end{eqnarray}
Notice that the last term disappears in standard cold inflation. In warm inflationary model, we define the slow roll parameters as follows:
\begin{eqnarray}
\varepsilon_{H} =\frac{\varepsilon}{1+Q}\,,\quad \eta_{H} = \frac{\eta}{1+Q}\,.
\label{slowroll}
\end{eqnarray}
Inflationary phase of the universe in warm inflation takes place when the slow-roll parameters satisfy the following conditions \cite{Hall:2003zp,Taylor:2000ze,Moss:2008yb}:
\begin{eqnarray}
\varepsilon \ll 1 + Q\,,\qquad \eta \ll 1 + Q\,,\qquad \beta \ll 1 + Q\,,\label{sloe}
\end{eqnarray}
where the condition on $\beta$ ensures that the variation of $\Gamma$ with respect to $\phi$ is slow enough. Compared to the cold scenario, the power spectrum of warm inflation gets modified and it is given in Refs.\cite{Graham2009,Bastero-Gil:2018uep,Hall:2003zp,Ramos:2013nsa,BasteroGil:2009ec,Taylor:2000ze,DeOliveira:2001he,Visinelli:2016rhn}
and it takes the form:
\begin{eqnarray}
P_{\cal R}(k) = \left( \frac{H_k^2}{2\pi\dot\phi_k}\right)^2\left( 1 + 2n_k +\left(\frac{T_k}{H_k}\right)\frac{2\sqrt{3}\,\pi\,Q_k}{\sqrt{3+4\pi\,Q_k}}\right)G(Q_k)\,,
\label{spectrum}
\end{eqnarray}
where the subscript $``k"$ signifies the time when the mode of cosmological perturbations with wavenumber $``k"$ leaves the horizon during inflation and $n = 1/\big( \exp{H/T} - 1 \big)$ is the Bose-Einstein distribution function. Additionally, the function $G(Q_k)$ encodes the coupling between the inflaton and the radiation in the heat bath leading to a growing mode in the fluctuations of the inflaton field. It is originally proposed in Ref.\cite{Graham2009} and its consequent implications can be found in Refs.\cite{BasteroGil:2011xd,BasteroGil:2009ec}.

This growth factor $G(Q_k)$ is dependent on the form of $\Gamma$ and is obtained numerically. As given in Refs.\cite{Benetti:2016jhf,Bastero-Gil:2018uep}, we see that for $\Gamma \propto T$:
\be
G(Q_k)_{\rm linear} = 1+0.0185Q^{2.315}_{k}+0.335Q^{1.364}_{k}
\,. \label{ga}
\ee
In this work, we consider a linear form of $G(Q_k)$ with $Q\gg 1$. Clearly, for small $Q$, i.e., $Q\ll 1$, the growth factor does not enhance the power spectrum. It is called the weak dissipation regime. However, for large $Q$, i.e., $Q\gg 1$, the growth factor significantly enhances the power spectrum. The latter is called the strong dissipation regime. The primordial tensor fluctuations of the metric give rise to a tensor power spectrum. It is the same form as that of cold inflation given
in Ref.\cite{Bartrum:2013fia} as
\be
P_{T}(k) = \frac{16}{\pi}\Big(\frac{H_{k}}{M_{p}}\Big)^{2}
\,. \label{PT}
\ee
The ratio of the tensor to the scalar power spectrum is expressed in terms of a parameter $r$ as
\be
r = \frac{P_{T}(k)}{P_{\cal R}(k)}\,. \label{r}
\ee

\section{Warm deformed $R^{2}$ scenario}\label{s4}
In the present analysis, we will consider warm inflation in the strong regime that the inflaton perturbations are non-trivially affected by the fluctuations of the thermal bath, and the amplitude of the spectrum may get a correction, generically called the ``growing mode”, depending on the value of the dissipative ratio. This was originally conducted by Graham and Moss \cite{Graham2009}. Since the solutions when $p=2$ are well known, hence we rewrite the potential by substituting $p/(p-1)=2(1-\delta)$ with $\delta\ll 1$. Therefore, we can use perturbation theory in the small parameter $\delta$. The resulting potential (\ref{Uep}) takes the form
\begin{eqnarray}\label{Uepdel}
U(\phi)=U_{0}e^{-2\sqrt{2/3}\kappa\phi}\Big(e^{\sqrt{2/3}\kappa\phi}-1\Big)^{2(1-\delta)}\,.
\end{eqnarray}
where
\begin{eqnarray}\label{Ueppdel}
U_{0}=\frac{1}{2\kappa^{2}}(p-1)p^{p/(1-p)}\lambda^{1/(1-p)}\quad{\rm with}\quad p=\frac{2 (\delta -1)}{2 \delta -1}\simeq 2+2\delta+{\cal O}(\delta^{2})\,.
\end{eqnarray}
From Eq.(\ref{SR-parameters}), we cam compute the slow-roll parameters to obtain
\begin{eqnarray}
\varepsilon &=& \frac{4 \left(\delta  e^{\frac{\sqrt{\frac{2}{3}} \phi }{M_{p}}}-1\right)^2}{3 \left(e^{\frac{\sqrt{\frac{2}{3}} \phi }{M_{p}}}-1\right)^2}\simeq\epsilon_{2}(\phi)\left(1-\delta\,e^{\frac{\sqrt{\frac{2}{3}} \phi }{M_{p}}}\right)\,,\\\eta &=& \frac{4 e^{\frac{\sqrt{\frac{2}{3}} \phi }{M_{p}}} \left(\delta  \left(2 \delta  e^{\frac{\sqrt{\frac{2}{3}} \phi }{M_{p}}}-3\right)-1\right)+8}{3 \left(e^{\frac{\sqrt{\frac{2}{3}} \phi }{M_{p}}}-1\right)^2}\simeq\eta_{2}(\phi)\left(1-\frac{3 \delta  e^{\frac{\sqrt{\frac{2}{3}} \phi }{M_{p}}}}{e^{\frac{\sqrt{\frac{2}{3}} \phi }{M_{p}}}-2}\right)\,,\\
\beta &=&\frac{4 \left(2 e^{\frac{\sqrt{\frac{2}{3}} \phi }{M_{p}}}-3\right) \left(\delta  e^{\frac{\sqrt{\frac{2}{3}} \phi }{M_{p}}}-1\right)}{15 \left(e^{\frac{\sqrt{\frac{2}{3}} \phi }{M_{p}}}-1\right)^2}\simeq\beta_{2}(\phi)  \left(1-\delta  e^{\frac{\sqrt{\frac{2}{3}} \phi }{M_{p}}}\right)\,,
\label{slow-roll-modelgen1}
\end{eqnarray}
where we have defined slow-roll parameters for $p=2$ as 
\begin{eqnarray}
\varepsilon_{2}(\phi) = \frac{4}{3 \left(e^{\frac{\sqrt{\frac{2}{3}} \phi }{M_{p}}}-1\right)^2}\,,\,\,\eta_{2}(\phi) = -\frac{4 \left(e^{\frac{\sqrt{\frac{2}{3}} \phi }{M_{p}}}-2\right)}{3 \left(e^{\frac{\sqrt{\frac{2}{3}} \phi }{M_{p}}}-1\right)^2}\,,\,\,
\beta_{2}(\phi) = -\frac{4 \left(2 e^{\frac{\sqrt{\frac{2}{3}} \phi }{M_{p}}}-3\right)}{15 \left(e^{\frac{\sqrt{\frac{2}{3}} \phi }{M_{p}}}-1\right)^2}\,,
\label{slow-roll-modelgen12}
\end{eqnarray}
We find $Q$ for the strong limit:
\begin{eqnarray}
Q \simeq \frac{\sqrt[5]{2} C_1}{3^{3/5}}\frac{\Bigg(\frac{U_0 e^{-\frac{\sqrt{6} \phi }{M_p}} \sqrt{\frac{U_0 \left(e^{\frac{\sqrt{\frac{2}{3}} \phi }{M_p}}-1\right)^2}{M_p^2}}}{C_1 C_r}\Bigg)}{\sqrt{\frac{U_0 e^{-\frac{2 \sqrt{\frac{2}{3}} \phi }{M_p}} \left(e^{\frac{\sqrt{\frac{2}{3}} \phi }{M_p}}-1\right){}^2}{M_p^2}}}\left(1-\delta  e^{\frac{\sqrt{\frac{2}{3}} \phi }{M_{p}}}\right)= Q_{2}(\phi) \left(1-\delta  e^{\frac{\sqrt{\frac{2}{3}} \phi }{M_{p}}}\right)\,,
\label{Q-strongE}
\end{eqnarray}
where
\begin{eqnarray}
Q_{2}(\phi) =\frac{\sqrt[5]{2} C_1}{3^{3/5}}\frac{\Bigg(\frac{U_0 e^{-\frac{\sqrt{6} \phi }{M_p}} \sqrt{\frac{U_0 \left(e^{\frac{\sqrt{\frac{2}{3}} \phi }{M_p}}-1\right)^2}{M_p^2}}}{C_1 C_r}\Bigg)}{\sqrt{\frac{U_0 e^{-\frac{2 \sqrt{\frac{2}{3}} \phi }{M_p}} \left(e^{\frac{\sqrt{\frac{2}{3}} \phi }{M_p}}-1\right)^2}{M_p^2}}}\,,
\label{Q-strongE2}
\end{eqnarray}

Our strategic analysis here is that we first solve the system for $\delta=0$, and then use perturbation theory in the small parameter $\delta$ and search
for a solution to this condition of the type
\begin{eqnarray}
\phi_{\rm end} =\phi^{\delta=0}_{\rm end}+\phi_{1}\delta\,.
\label{sodel0}
\end{eqnarray}
For $\delta=0$, we consider 
\begin{eqnarray}
\varepsilon_{2}(\phi) &=& \frac{4}{3 \left(e^{\frac{\sqrt{\frac{2}{3}} \phi }{M_{p}}}-1\right)^2},\\
Q_{2}(\phi) &=&\frac{\sqrt[5]{2} C_1}{3^{3/5}}\frac{\Bigg(\frac{U_0 e^{-\frac{\sqrt{6} \phi }{M_p}} \sqrt{\frac{U_0 \left(e^{\frac{\sqrt{\frac{2}{3}} \phi }{M_p}}-1\right)^2}{M_p^2}}}{C_1 C_r}\Bigg)}{\sqrt{\frac{U_0 e^{-\frac{2 \sqrt{\frac{2}{3}} \phi }{M_p}} \left(e^{\frac{\sqrt{\frac{2}{3}} \phi }{M_p}}-1\right)^2}{M_p^2}}}\,,
\label{del-0}
\end{eqnarray}
When inflation ends, one finds from Eq.(\ref{sloe}) using a condition $\varepsilon_{\rm end} \approx Q_{\rm end}$ with $\delta=0$:
\begin{eqnarray}
\frac{4}{3 \left(e^{\frac{\sqrt{\frac{2}{3}} \phi^{\delta=0}_{\rm end} }{M_{p}}}-1\right)^2} \approx \frac{\sqrt[5]{2} C_1}{3^{3/5}}\frac{\Bigg(\frac{U_0 e^{-\frac{\sqrt{6} \phi^{\delta=0}_{\rm end} }{M_p}} \sqrt{\frac{U_0 \left(e^{\frac{\sqrt{\frac{2}{3}} \phi^{\delta=0}_{\rm end} }{M_p}}-1\right)^2}{M_p^2}}}{C_1 C_r}\Bigg)}{\sqrt{\frac{U_0 e^{-\frac{2 \sqrt{\frac{2}{3}} \phi^{\delta=0}_{\rm end} }{M_p}} \left(e^{\frac{\sqrt{\frac{2}{3}} \phi^{\delta=0}_{\rm end} }{M_p}}-1\right)^2}{M_p^2}}}\,,\label{endE0}
\end{eqnarray}
Apparently, the above equation can be analytically solved to obtain exact solutions. To this end, we can solve Eq.(\ref{endE0}) to obtain a value of the inflaton field at the end of inflation to yield
\begin{eqnarray}
\phi^{\delta=0}_{\rm end} \approx \sqrt{\frac{3}{2}}\frac{1}{8}M_{p}\log{\Big(\frac{2^9 C_{r} U_{0}}{3^2 C_{1}M_{p}^4}\Big)},\label{so0}
\end{eqnarray}
where a large field approximation has been implemented by assuming $e^{\sqrt{\frac{2}{3}}\phi_{\rm end}/M_p}\pm 1 \approx e^{\sqrt{\frac{2}{3}}\phi_{\rm end}/M_p}$. Substituting a solution (\ref{so0}) into Eq.(\ref{sodel0}) and then applying to Eq.(\ref{endE0}), we find for $\phi_{1}$:
\begin{eqnarray}
\phi_{1}\simeq\frac{1}{5} 2^{5/8} \sqrt[4]{3} M_{p} \sqrt[8]{\frac{C_{r} U_{0}}{C_{1} M_{p}^4}}\,.\label{p1}
\end{eqnarray}
Therefore, the solution reads
\begin{eqnarray}
\phi_{\rm end}=\phi^{\delta=0}_{\rm end}+\phi_{1}\delta \simeq \sqrt{\frac{3}{2}}\frac{1}{8}M_{p}\log{\Big(\frac{2^9 C_{r} U_{0}}{3^2 C_{1}M_{p}^4}\Big)}+\frac{1}{5} 2^{5/8} \sqrt[4]{3} M_{p} \sqrt[8]{\frac{C_{r} U_{0}}{C_{1} M_{p}^4}}\delta\,.\label{sosp1}
\end{eqnarray}
Taking $\delta=0$, we simply obtain the results of $R^{2}$-type warm inflation, see Ref.\cite{Samart:2021eph}. Moreover, the inflaton field at the Hubble horizon crossing in the strong regime, $\phi_{N}$, can be determined using the perturbation trick. For the number of e-folding, we see that 
\begin{eqnarray}
N &=& \frac{1}{M_p^2}\int_{\phi_{\rm end}}^{\phi_{\rm ini}}\frac{Q\,U}{U'}\,d\phi
\nonumber\\
&=& \frac{5C_{1}\Bigg(8 e^{\frac{\sqrt{\frac{2}{3}} \phi }{M_{p}}}+12\Bigg)\Bigg(\frac{U_{0} \sqrt{\frac{U_{0}}{M_{p}^2}} e^{-\frac{2 \sqrt{\frac{2}{3}} \phi }{M_{p}}}}{C_{1} C_{r}}\Bigg)}{16\, 2^{4/5} 3^{3/5} \sqrt{\frac{U_{0}}{M_{p}^2}}}\Bigg(1+\delta  e^{\frac{\sqrt{\frac{2}{3}} \phi }{M_{p}}}\Bigg)\Bigg|^{\phi_{\rm ini}}_{\phi_{\rm end}}\,.
\label{N-strongE}
\end{eqnarray}
We search for the solution of the type:
\begin{eqnarray}
\phi_{N} =\phi^{\delta=0}_{N}+\phi_{2}\delta\,.
\label{sodelN}
\end{eqnarray}
Consider Eq.(\ref{N-strongE}) using $\delta=0$, we find for $\phi^{\delta=0}_{N}$: \begin{eqnarray}
\phi^{\delta=0}_{N}=\sqrt{\frac{3}{2}}\frac{1}{3}M_{p}\log{\Big(\frac{2^5\,2^4\,3^3 C_{r} N^5 U_{0}}{5^5\,C_{1}^4 M_{p}^4}\Big)}\,.
\label{sodelN1}
\end{eqnarray}
Substituting Eq.(\ref{sodelN1}) into Eq.(\ref{sodelN}), and applying back to Eq.(\ref{N-strongE}), we then solve to obtain \begin{eqnarray}
\phi_{N}&=&\phi^{\delta=0}_{N}+\phi_{2}\delta\nonumber\\&=&\sqrt{\frac{3}{2}}\frac{1}{3}M_{p}\log{\Big(\frac{2^5\,2^4\,3^3 C_{r} N^5 U_{0}}{5^5\,C_{1}^4 M_{p}^4}\Big)}\nonumber\\&+&\frac{4 \sqrt{6} M_p}{5 \left(5\ 5^{2/3}-24 \sqrt[3]{\frac{N^5 U_0 C_r}{C_1^4 M_p^4}}\right)}\sqrt[3]{\frac{N^5 U_0 C_r}{C_1^4 M_p^4}} \left(9 \sqrt[3]{5} \sqrt[3]{\frac{N^5 U_0 C_r}{C_1^4 M_p^4}}-25\right)\delta\,.
\label{sodelN1}
\end{eqnarray}
As done above, we therefore can re-write the slow-roll parameters in terms of the number of {\it e}-foldings, $N$, by using large field approximation in the strong $Q$ limit and then we find
\begin{eqnarray}
\varepsilon &\approx& \frac{125 \sqrt[3]{5} C_1^4 M_p^4 \sqrt[3]{\frac{N^5 U_0 C_r}{C_1^4 M_p^4}}}{432 N^5 U_0 C_r}-\frac{25 \delta  \left(9\ 5^{2/3} N^5 U_0 C_r-25 \sqrt[3]{5} C_1^4 M_p^4 \left(\frac{N^5 U_0 C_r}{C_1^4 M_p^4}\right){}^{2/3}\right)}{27 N^5 U_0 C_r \left(5\ 5^{2/3}-24 \sqrt[3]{\frac{N^5 U_0 C_r}{C_1^4 M_p^4}}\right)}\,,\\\eta &\approx& -\frac{5\ 5^{2/3} C_1^4 M_p^4 \left(\frac{N^5 U_0 C_r}{C_1^4 M_p^4}\right){}^{2/3}}{18 N^5 U_0 C_r}-\frac{20 \delta  \left(5\ 5^{2/3}-9 \sqrt[3]{\frac{N^5 U_0 C_r}{C_1^4 M_p^4}}\right)}{9 \left(5\ 5^{2/3}-24 \sqrt[3]{\frac{N^5 U_0 C_r}{C_1^4 M_p^4}}\right)}\,,\\\beta &\approx& -\frac{5^{2/3} C_1^4 M_p^4 \left(\frac{N^5 U_0 C_r}{C_1^4 M_p^4}\right){}^{2/3}}{9 N^5 U_0 C_r}-\frac{8 \delta  \left(5\ 5^{2/3}-9 \sqrt[3]{\frac{N^5 U_0 C_r}{C_1^4 M_p^4}}\right)}{9 \left(5\ 5^{2/3}-24 \sqrt[3]{\frac{N^5 U_0 C_r}{C_1^4 M_p^4}}\right)}\,.
\label{slow-roll-modelEN}
\end{eqnarray}
Moreover, we can write $Q$ for the strong limit in terms of $N$ as 
\begin{eqnarray}
Q_{N} &\simeq& \frac{5^{2/3} C_1^5 M_p^6 \sqrt{\frac{U_0}{M_p^2}} \sqrt[5]{\frac{U_0 \sqrt{\frac{U_0}{M_p^2}}}{C_1 C_r}} \left(\frac{N^5 U_0 C_r}{C_1^4 M_p^4}\right){}^{13/15}}{6 N^5 U_0^2 C_r}\nonumber\\&-&\frac{8 C_1 \delta  M_p^2 \sqrt{\frac{U_0}{M_p^2}} \sqrt[5]{\frac{U_0 \sqrt{\frac{U_0}{M_p^2}}}{C_1 C_r}} \left(63 \left(\frac{N^5 U_0 C_r}{C_1^4 M_p^4}\right){}^{8/15}-10\ 5^{2/3} \sqrt[5]{\frac{N^5 U_0 C_r}{C_1^4 M_p^4}}\right)}{15 U_0 \left(24 \sqrt[3]{\frac{N^5 U_0 C_r}{C_1^4 M_p^4}}-5\ 5^{2/3}\right)}\,.
\label{Q-strongEN}
\end{eqnarray}
It is noticed that for a large field approximation the results given above do depend on a small number, $\delta$, as expected. When setting $\delta=0$, we have the results of warm scenario for $R^{2}$-type inflation, see for instance Ref.\cite{Samart:2021eph} for Higgs-Starobinky inflation. Since the energy density during inflation is predominated by its potential of the inflaton field. Therefore we can write the Einstein equation as
\be
H^{2} = \frac{8\pi}{3}\frac{U}{M^{2}_{p}}=\frac{8\pi}{3}\frac{U_{0}}{M^{2}_{p}} e^{-\frac{2\sqrt{\frac{2}{3}} \phi_{k}}{M_p}}\Big( e^{\frac{\sqrt{\frac{2}{3}} \phi_{k}}{M_p}}-1\Big)^{2(1-\delta)}\,,\label{H2}
\ee
Using the above relation, we can write Eq.(\ref{eominfsl}) for our model as
\begin{eqnarray}
\dot{\phi} &\simeq&  -\frac{U_\phi}{3H(1+Q)}\simeq -\frac{M_p \sqrt{\frac{U_0}{M_p^2}}}{3 \sqrt{\pi } (Q+1)}(e^{\frac{-\sqrt{\frac{2}{3}} \phi }{M_p}}+\delta)\,,\label{phidot}
\end{eqnarray}
Then combining these two quantities, we end up with to the first order of $\delta$:
\begin{eqnarray}
\frac{H^{2}_{k}}{2\pi\dot{\phi}_{k}}\simeq -\frac{4\sqrt{\pi} (Q+1)}{M_p}\sqrt{\frac{U_0}{M_p^2}} e^{\frac{\sqrt{\frac{2}{3}} \phi_{k}}{M_p}}(1+e^{\frac{\sqrt{\frac{2}{3}} \phi_{k} }{M_p}}\delta)\,.\label{1st}
\end{eqnarray}
On substituting Eq.(\ref{phidot}) in the energy density of radiation given in Eq.(\ref{eomrad}), we obtain the temperature of the thermal bath as
\begin{eqnarray}
T_{k}= \Big(\frac{3Q\dot{\phi}^{2}}{4\,C_{r}}\Big)^{1/4}\simeq \frac{1}{\sqrt{2} \sqrt[4]{3 \pi }}\Bigg(\frac{Q U_0 e^{-\frac{2 \sqrt{\frac{2}{3}} \phi_{k}}{M_p}}}{C_{r} (Q+1)^2}\Bigg)^{1/4}(1-\delta\,e^{\frac{\sqrt{\frac{2}{3}} \phi_{k}}{M_p}}/2)\,,\label{Tk}
\end{eqnarray}
with $C_{r}=\pi^{2}g_{*}/30$ where $g_{*}$ is the number of relativistic degrees of freedom during warm inflation. Regarding Ref.\cite{Arya:2018sgw}, we can take $g_{*}\approx 200$. Then we can combine the above result with $H$ from Eq.(\ref{H2}) to obtain the factor $T/H$ to yield
\begin{eqnarray}
\frac{T_{k}}{H_{k}}\simeq \frac{\sqrt[4]{3}}{4 \pi ^{3/4} \sqrt{\frac{U_0}{M_p^2}}}\left(\frac{Q U_0 e^{-\frac{2 \sqrt{\frac{2}{3}} \phi }{M_p}}}{C_{r} (Q+1)^2}\right)^{1/4}(1-e^{\frac{\sqrt{\frac{2}{3}} \phi_{k} }{M_p}}\delta/2)\,.\label{TkH}
\end{eqnarray}
Since the dissipation parameter is defined as $Q=\frac{\Gamma}{3H}$ for model of warm inflation, we consider $\Gamma= C_1 T$. After substituting this form of $\Gamma$ we obtain $T= \frac{3HQ}{C_1}$. We equate this with Eq. (\ref{TkH}) to obtain
 \begin{eqnarray}
\frac{\phi_{k}}{M_p}\simeq\sqrt{\frac{3}{2}}\log\Bigg({\cal A}\Bigg(\Bigg(\sqrt{3} C_1^2 \delta  \sqrt{\frac{2 Q U_0}{C_{r} (Q+1)^2}}+\frac{72 \pi ^{3/2} Q^2 U_0}{M_p^2}\Bigg)^{1/2}-12 \pi^{3/4} Q \sqrt{\frac{U_0}{M_p^2}}\Bigg)^2\Bigg)\,,
 \label{phik}
 \end{eqnarray}
where
 \begin{eqnarray}
{\cal A}=\frac{1}{C_1^2 \delta^2 \sqrt{\frac{3 Q U_0}{C_{r} (Q+1)^2}}}\,,
\label{def}
\end{eqnarray}
and we have assumed a large field approximation to write Eq.(\ref{phik}). For the dissipation–dominated regime, the dissipation rate, $\Gamma$, is much greater than the
expansion rate, i.e., $Q\gg 1$. In this case, the evolution of the inflaton field during this phase can be approximately obtained. This allows us to write the energy density of the radiation field as
 \begin{equation}
\rho_{R}(k)\simeq \frac{1}{12Q_{k}}\left(\frac{U_{\phi_{k}}}{H_{k}}\right)^{2}=\frac{U_0}{12 \pi Q_{k}}e^{-\frac{2 \sqrt{\frac{2}{3}} \phi_{k}}{M_p}}\left(1-2 \delta e^{\frac{\sqrt{\frac{2}{3}} \phi_{k}}{M_p}}\right)\,,
 \label{rhor}
 \end{equation}	
Substituting results given in Eq.(\ref{phik}) in Eqs.(\ref{TkH}) and (\ref{1st}), we can express $P_\mathcal{R}(k)$ in terms of variables $Q_k,\,C_{r}$ and $C_1$. Also, from its definition in Eq. (\ref{slowroll}), the slow roll parameter can be written as
\be
\varepsilon_{H} &=& \frac{\varepsilon}{(1+Q_k)}\simeq \frac{1}{(1+Q_k)}\frac{4}{3 \left(e^{\frac{\sqrt{\frac{2}{3}} \phi_{k}}{M_{p}}}-1\right)^2}\left(1-2 \delta  e^{\frac{\sqrt{\frac{2}{3}} \phi_{k}}{M_{p}}}\right)\,,\\\eta_{H}&=& \frac{\eta}{(1+Q_k)}\simeq \frac{1}{(1+Q_k)}\frac{8 - 4 (3 \delta +1)\,e^{\frac{\sqrt{\frac{2}{3}} \phi_{k}}{M_p}}}{3 \left(e^{\frac{\sqrt{\frac{2}{3}} \phi_{k}}{M_p}}-1\right)^2}.
\label{eph1}
\ee
From Eq.(\ref{H2}), we can write
\be
\left(\frac{H}{M_p}\right)^{2} =\frac{8\pi}{3}\frac{U_{0}}{M^{4}_{p}}e^{-2\sqrt{2/3}\kappa\phi}\Big(e^{\sqrt{2/3}\kappa\phi}-1\Big)^{2(1-\delta)}\,.\label{H22}
\ee
Using Eq.({\ref{H22}}), the tensor power spectrum for this model is evaluated as
\be
P_T(k)= \frac{16}{\pi}\left(\frac{H_k}{M_{Pl}}\right)^2\simeq \frac{128}{3}\frac{U_{0}}{M^{4}_{p}}e^{-\frac{2\sqrt{\frac{2}{3}} \phi_{k}}{M_p}}\Big( e^{\frac{\sqrt{\frac{2}{3}} \phi_{k}}{M_p}}-1\Big)^{2(1-\delta)}.
\ee
Note here that we can use Eq.(\ref{phik}) and can express $P_T(k)$ in terms of model parameters.

\section{Background dynamics}
In this section, a detailed analysis of the background dynamics, considering the evolution of the radiation energy density, $\rho_{R}$, and the quantities that are important for warm inflation, e.g., $\phi/M_{p}$, $Q$, $T/H$, and so on, would be interesting to be examined. We start in this section studying how the dissipation parameter, $Q$, evolves with the number of efolds, $N$.
\begin{eqnarray}
\frac{dQ}{dN}&\simeq& \frac{13\ 5^{2/3} C_1 M_{p}^2 \sqrt{\frac{U_0}{M_{p}^2}} \sqrt[5]{\frac{U_0 \sqrt{\frac{U_0}{M_{p}^2}}}{C_1 C_r}}}{18 N U_0 \left(\frac{N^5 U_0 C_r}{C_1^4 M_{p}^4}\right)^{2/15}}\nonumber\\&+&\frac{64 \delta  N^4 C_r \sqrt{\frac{U_0}{M_{p}^2}} \sqrt[5]{\frac{U_0 \sqrt{\frac{U_0}{M_{p}^2}}}{C_1 C_r}} \left(63 \left(\frac{N^5 U_0 C_r}{C_1^4 M_{p}^4}\right)^{8/15}-10\ 5^{2/3} \sqrt[5]{\frac{N^5 U_0 C_r}{C_1^4 M_{p}^4}}\right)}{3 C_1^3 M_{p}^2 \left(\frac{N^5 U_0 C_r}{C_1^4 M_{p}^4}\right)^{2/3} \left(24 \sqrt[3]{\frac{N^5 U_0 C_r}{C_1^4 M_{p}^4}}-5\ 5^{2/3}\right){}^2}\nonumber\\&-&\frac{8 C_1 \delta M_{p}^2 \sqrt{\frac{U_0}{M_{p}^2}} \sqrt[5]{\frac{U_0 \sqrt{\frac{U_0}{M_{p}^2}}}{C_1 C_r}} \left(\frac{168 N^4 U_0 C_r}{C_1^4 M_{p}^4 \left(\frac{N^5 U_0 C_r}{C_1^4 M_{p}^4}\right){}^{7/15}}-\frac{10\ 5^{2/3} N^4 U_0 C_r}{C_1^4 M_{p}^4 \left(\frac{N^5 U_0 C_r}{C_1^4 M_{p}^4}\right)^{4/5}}\right)}{15 U_0 \left(24 \sqrt[3]{\frac{N^5 U_0 C_r}{C_1^4 M_{p}^4}}-5\ 5^{2/3}\right)}\nonumber\\&-&\frac{5\ 5^{2/3} C_1^5 M_{p}^6 \sqrt{\frac{U_0}{M_{p}^2}} \sqrt[5]{\frac{U_0 \sqrt{\frac{U_0}{M_{p}^2}}}{C_1 C_r}} \left(\frac{N^5 U_0 C_r}{C_1^4 M_{p}^4}\right){}^{13/15}}{6 N^6 U_0^2 C_r}\,.
\end{eqnarray}
For the inflaton field, we have
\begin{eqnarray}
\frac{d\phi/M_{p}}{dN}&\simeq&\frac{12 \sqrt[3]{5} \sqrt{6} \delta  N^4 U_0 C_r}{C_1^4 \text{Mp}^4 \sqrt[3]{\frac{N^5 U_0 C_r}{C_1^4 \text{Mp}^4}} \left(5\ 5^{2/3}-24 \sqrt[3]{\frac{N^5 U_0 C_r}{C_1^4 \text{Mp}^4}}\right)}\nonumber\\&+&\frac{4 \sqrt{\frac{2}{3}} \delta  N^4 U_0 C_r \left(9 \sqrt[3]{5} \sqrt[3]{\frac{N^5 U_0 C_r}{C_1^4 \text{Mp}^4}}-25\right)}{C_1^4 \text{Mp}^4 \left(\frac{N^5 U_0 C_r}{C_1^4 \text{Mp}^4}\right)^{2/3} \left(5\ 5^{2/3}-24 \sqrt[3]{\frac{N^5 U_0 C_r}{C_1^4 \text{Mp}^4}}\right)}\nonumber\\&+&\frac{32 \sqrt{6} \delta  N^4 U_0 C_r \left(9 \sqrt[3]{5} \sqrt[3]{\frac{N^5 U_0 C_r}{C_1^4 \text{Mp}^4}}-25\right)}{C_1^4 \text{Mp}^4 \sqrt[3]{\frac{N^5 U_0 C_r}{C_1^4 \text{Mp}^4}} \left(5\ 5^{2/3}-24 \sqrt[3]{\frac{N^5 U_0 C_r}{C_1^4 \text{Mp}^4}}\right){}^2}+\frac{5}{\sqrt{6} N}\,.
\end{eqnarray}
and
\begin{eqnarray}
T/H&\simeq& \frac{5}{8 \sqrt[4]{2} \pi^{3/4} \sqrt{\frac{U_0}{M_{p}^2}}}{\cal B}+{\cal C}^{-1}\Bigg( \left(\frac{2}{\pi }\right)^{3/4} \delta M_{p}^2 U_0 \Bigg(27 \sqrt[3]{5} N^5 C_r \sqrt{\frac{U_0}{M_{p}^2}} \left(\frac{N^5 U_0 C_r}{C_1^4 M_{p}^4}\right)^{2/3}\nonumber\\&&\quad\quad\quad\quad\quad\quad\quad\quad\quad\quad+5\ 5^{2/3} C_1 N^5 C_r \sqrt[5]{\frac{U_0 \sqrt{\frac{U_0}{M_{p}^2}}}{C_1 C_r}} \sqrt[5]{\frac{N^5 U_0 C_r}{C_1^4 M_{p}^4}}\nonumber\\&&\quad\quad\quad\quad\quad\quad\quad\quad\quad\quad-9 C_1 N^5 C_r \sqrt[5]{\frac{U_0 \sqrt{\frac{U_0}{M_{p}^2}}}{C_1 C_r}} \left(\frac{N^5 U_0 C_r}{C_1^4 M_{p}^4}\right)^{8/15}\Bigg){\cal B}\Bigg)\,.
\end{eqnarray}
where
\begin{eqnarray}
{\cal B}&\equiv&\left(\frac{C_1^5 M_{p}^6 N^5 U_0^3 \sqrt{\frac{U_0}{M_{p}^2}} \sqrt[5]{\frac{U_0 \sqrt{\frac{U_0}{M_{p}^2}}}{C_1 C_r}} \sqrt[5]{\frac{N^5 U_0 C_r}{C_1^4 M_{p}^4}}}{\left(5^{2/3} C_1^5 M_{p}^6 \sqrt{\frac{U_0}{M_{p}^2}} \sqrt[5]{\frac{U_0 \sqrt{\frac{U_0}{M_{p}^2}}}{C_1 C_r}} \left(\frac{N^5 U_0 C_r}{C_1^4 M_{p}^4}\right)^{13/15}+6 N^5 U_0^2 C_r\right)^2}\right)^{1/4}\nonumber\\{\cal C}&=&\left(5\ 5^{2/3}-24 \sqrt[3]{\frac{N^5 U_0 C_r}{C_1^4 M_{p}^4}}\right) \left(5^{2/3} C_1^5 M_{p}^6 \sqrt{\frac{U_0}{M_{p}^2}} \sqrt[5]{\frac{U_0 \sqrt{\frac{U_0}{M_{p}^2}}}{C_1 C_r}} \left(\frac{N^5 U_0 C_r}{C_1^4 M_{p}^4}\right){}^{13/15}+6 N^5 U_0^2 C_r\right)\,.
\end{eqnarray}
For the energy density of radiation, we find
\begin{eqnarray}
\frac{d\rho_{R}}{dN}&\simeq&-\frac{20\ 5^{2/3} \delta  C_r \left(\frac{U_0 \sqrt{\frac{U_0}{M_{p}^2}}}{C_1 C_r}\right)^{4/5} \left(9 \sqrt[3]{5} N^5 U_0 C_r \left(\frac{N^5 U_0 C_r}{C_1^4 M_{p}^4}\right)^{2/15}-25 C_1^4 M_{p}^4 \left(\frac{N^5 U_0 C_r}{C_1^4 M_{p}^4}\right){}^{4/5}\right)}{9 \pi  C_1^4 M_{p}^4 N \left(\frac{N^5 U_0 C_r}{C_1^4 M_{p}^4}\right)^{2/3} \left(5\ 5^{2/3}-24 \sqrt[3]{\frac{N^5 U_0 C_r}{C_1^4 M_{p}^4}}\right)^2}\nonumber\\&+&\frac{175\ 5^{2/3} C_r \left(\frac{U_0 \sqrt{\frac{U_0}{M_{p}^2}}}{C_1 C_r}\right)^{4/5}}{3456 \pi  N \left(\frac{N^5 U_0 C_r}{C_1^4 M_{p}^4}\right)^{8/15}}-\frac{125\ 5^{2/3} C_1^4 M_{p}^4 \left(\frac{U_0 \sqrt{\frac{U_0}{M_{p}^2}}}{C_1 C_r}\right)^{4/5} \left(\frac{N^5 U_0 C_r}{C_1^4 M_{p}^4}\right)^{7/15}}{1152 \pi  N^6 U_0}\\&+&\frac{5\ 5^{2/3} \delta  \left(\frac{U_0 \sqrt{\frac{U_0}{M_{p}^2}}}{C_1 C_r}\right)^{4/5} \left(9 \sqrt[3]{5} N^5 U_0 C_r \left(\frac{N^5 U_0 C_r}{C_1^4 M_{p}^4}\right)^{2/15}-25 C_1^4 M_{p}^4 \left(\frac{N^5 U_0 C_r}{C_1^4 M_{p}^4}\right)^{4/5}\right)}{18 \pi  N^6 U_0 \left(5\ 5^{2/3}-24 \sqrt[3]{\frac{N^5 U_0 C_r}{C_1^4 M_{p}^4}}\right)}\nonumber\\&-&\frac{5^{2/3} \delta  \left(\frac{U_0 \sqrt{\frac{U_0}{M_{p}^2}}}{C_1 C_r}\right)^{4/5} \left(\frac{6 \sqrt[3]{5} N^9 U_0^2 C_r^2}{C_1^4 M_{p}^4 \left(\frac{N^5 U_0 C_r}{C_1^4 M_{p}^4}\right){}^{13/15}}+45 \sqrt[3]{5} N^4 U_0 C_r \left(\frac{N^5 U_0 C_r}{C_1^4 M_{p}^4}\right)^{2/15}-\frac{100 N^4 U_0 C_r}{\sqrt[5]{\frac{N^5 U_0 C_r}{C_1^4 M_{p}^4}}}\right)}{18 \pi  N^5 U_0 \left(5\ 5^{2/3}-24 \sqrt[3]{\frac{N^5 U_0 C_r}{C_1^4 M_{p}^4}}\right)}\,.\nonumber
\end{eqnarray}

\begin{figure}[!h]	
	\includegraphics[width=7cm]{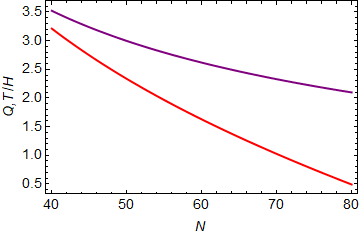}
	\includegraphics[width=7cm]{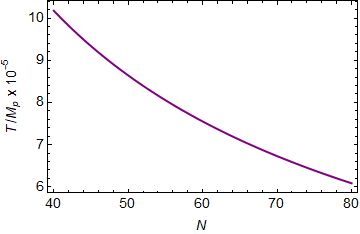}
	\includegraphics[width=7cm]{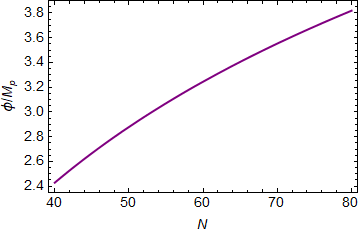}
	\includegraphics[width=7cm]{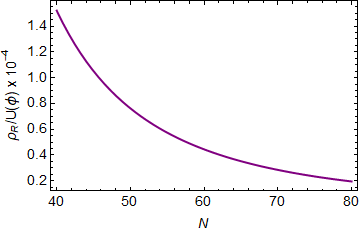}
	\centering
	\caption{We display the dynamical evolution in warm inflation with a potential given in Eq.(\ref{Uepdel}). The behaviour of the dissipation
parameter $Q$ (purple line), the ratio $T/H$ (red line), temperature of
the Universe $T$ (in units of $M_{p}$), the homogeneous inflaton field $\phi$ (in units of $M_{p}$), the energy density in $\phi$ is shown as a function of the number of efolds $N$ with the dissipation coefficient $\Gamma= C_{1} T$. To generate this plot, we take $C_{1}=0.3,\,C_{r}=70,\,\delta=0.02$ and $U_{0}=10^{-10}\,M^{4}_{p}$}
	\label{stplot111}
\end{figure}
We display the background dynamics by considering the evolution of the radiation energy density, $Q,\,T/H,\,T/M_{p}$ and $\phi/M_{p}$ in Fig.(\ref{stplot111}). We illustrate the evolution of the different dynamical quantities in the deformed $R^{2}$ model, obtained numerically for an example with $C_{r}=70,\,C_{1}=0.3,\,\delta=0.02$, and $U_{0}=10^{-10}\,M^{4}_{p}$.

\section{Confrontation with Planck 2018 data}\label{s5}
\begin{figure}[!h]	
	\includegraphics[width=10cm]{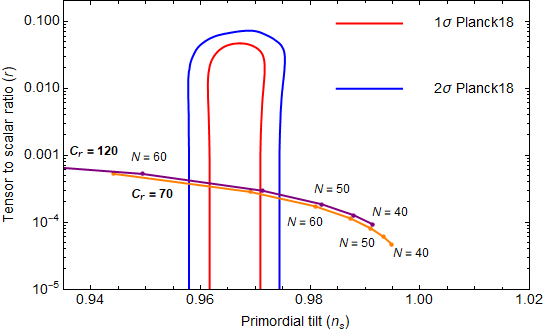}
	\centering
	\caption{We compare the theoretical predictions of $(r,\,n_s)$ in the strong limit $Q\gg 1$ for $R^{p}$ model. We consider a linear form of the growing mode function $G(Q_N)$. For the plots, we have used $U_{0}\simeq 10^{-10}\,M^{4}_{p},\,C_{1}=0.30,\,\delta=0.02$ and $C_{r}=70$ (orange line), and $C_{r}=120$ (purple line). We consider theoretical predictions of $(r,\,n_s)$ for different values
of $N$ with Planck’18 results for TT, TE, EE, +lowE+lensing+BK15+BAO.}
	\label{stplotG1}
\end{figure}

In this section, the inflation potentials can be constrained using the COBE normalization condition \cite{Bezrukov:2008ut}. This can be used to fix the parameters of the models in the present analysis. From Planck 2018 data, the inflaton potential must be normalized by the slow-roll parameter, $\epsilon$ and satisfied the following relation at the horizon crossing $\phi=\phi_{N}$ in order to produce the observed amplitude of the cosmological density perturbation ($A_{s}$):
\begin{eqnarray}
\frac{U(\phi_N)}{\varepsilon(\phi_N)} \simeq (0.0276\,M_{p})^{4}\,.
\label{Cobe-strong}
\end{eqnarray}
Taking the potential Eq.(\ref{Uepdel}) and the first slow-roll parameter $\varepsilon$ given in Eq.(\ref{slow-roll-modelgen1}), and substituting $\phi_{N}$ given in Eq.(\ref{sodelN1}), we approximately find that
\begin{eqnarray}
U_{0}\simeq \frac{1.2\times 10^{-4} C_1^{8/5} M_p^4}{N^2 C_r^{2/5}}+\frac{7.2\times 10^{-6} C_1^{4/5} M_p^4}{N \sqrt[5]{C_r}}\delta\,.
\end{eqnarray}
As of the primordial power spectrum for all the models written in terms of $Q$,\,$\lambda$,\,and $C_{1}$, we can demonstrate how the the power spectrum does depend on the scale. The spectral index of the primordial power spectrum is defined as
\begin{eqnarray}
n_{s}-1=\frac{d \ln P_{\cal R}(k)}{d\ln (k/k_{p})}=\frac{d\ln P_{\cal R}(k)}{dQ}\frac{dQ}{dN}\frac{dN}{dx}\Bigg|_{k=k_{p}}\,,\label{ns1}
\end{eqnarray}
where $x=\ln(x/x_{p})$ and $k_{p}$ corresponds to the pivot scale. From a definition of $N$, it is rather straightforward to show that \cite{Arya:2018sgw}
\begin{eqnarray}
\frac{dN}{dx}=-\frac{1}{1-\varepsilon_{H}}\,.
\end{eqnarray}
Now we compute $r$ and $n_{s}$ using Eq.(\ref{r}) and Eq.(\ref{ns1}) for a linear form of the growing mode function $G(Q)$ given in Eq.(\ref{ga}). Note that $r$ and $n_{s}$ are approximately given in Refs. \cite{Bastero-Gil:2018uep,BasteroGil:2009ec,Benetti:2016jhf}. We show the predictions of deformed $R^{p}$ gravity in Fig.(\ref{stplotG1}) where we have used two values of $C_{r}=70,\,120$. We have also found that if dissipation is already strong at horizon
crossing, $Q\gg 1$, the spectrum becomes more blue-tilted. This is due to the coupling between inflaton and radiation fluctuations. This behavior was noticed so far in Refs.\cite{Benetti:2016jhf,Bastero-Gil:2016qru}.

\section{Conclusion}\label{conclud}
In this work, we have investigated warm inflationary model in the context of a deformation of $R^{2}$ gravity which is coupled to radiation through a dissipation term. We start considering $R^{p}$ setup and assume $p=2(1+\delta)$ with $\delta\ll 1$ so that we can simply use the perturbation method. Particularly, our results covered simple warm $R^{2}$ inflation when setting $\delta=0$. We have demonstrated detailed derivations of the potentials in the Einstein frame, and derived relevant parameters in the warm $R^{p}$ inflation using the slow-roll approximation. Concretely, we have particularly considered a dissipation parameter of the form $\Gamma = C_{1}T$ with $C_1$ being a coupling parameter and have focused only on the strong regime of which the interaction between inflaton and radiation fluid has been taken into account. 

In this work, we have also taken into account a detailed analysis of the background dynamics, considering the evolution of the radiation energy density, $\rho_{R}$, and the quantities that are important for warm inflation, e.g., $Q,\,T/H,\,T/M_{p}$ and $\phi/M_{p}$. To confront the results with the data, we have computed inflationary observables and have constrained the parameters of our model using current Planck 2018 data. We have compared the theoretical predictions of $(r,\,n_{s})$ in the strong limit for the model with Planck’18 results. With proper choices of parameters, we have demonstrated that the predictions are in good agreement with Planck 2018 data \cite{Planck:2018jri}. Additionally, the potential scale $U_{0}$ of the models were constrained using the COBE normalization condition. It is worth noting that our scenario on warn deformed $R^{2}$ inflation may be possibly linked to the
marginally deformed Starobinsky model \cite{Codello:2014sua} dictating the trace-log quantum corrections. However, the deformation can be tested by current and future experimental
results and constitutes a sensible generalization
of the original (warm) Starobinsky scenario.

We should stress here that other forms of dissipation coefficient considered in the literature might also be
relevant to be considered. For example, a dissipation coefficient with a cubic dependence on the temperature given by $\Gamma=C_{\phi}T^{3}/\phi^{2}$ was studied in Refs.\cite{Berera:2008ar,Bastero-Gil:2011rva,Bastero-Gil:2010dgy,Berghaus:2019whh,Laine:2021ego,Motaharfar:2021egj}, while the high temperature regime with $\Gamma\propto T$ was found in Refs.\cite{Berera:2008ar,Moss:2008yb,Panotopoulos:2015qwa,Bastero-Gil:2016qru}. Additionally, for the case in which a dissipation coefficient depends only on the scalar field with $\Gamma\propto \phi^{-1}$ was first considered in warm inflation in Ref.\cite{deOliveira:1997jt}. However, based on the present analysis, analytical solutions of deformed $R^{2}$ gravity can not be obtained for those of the dissipation forms. This requires the numerical computations. We will leave them for future investigation.

\acknowledgments
P. Channuie acknowledged the Mid-Career Research Grant 2020 from National Research Council of Thailand (NRCT5-RSA63019-03).


\begin{thebibliography}{99}


\bibitem{Starobinsky:1980te} 
  A.~A.~Starobinsky,
  Phys.\ Lett.\ B {\bf 91}, 99 (1980).
  
\bibitem{Sato:1980yn}
K.~Sato,
Mon. Not. Roy. Astron. Soc. \textbf{195} (1981), 467-479
NORDITA-80-29.

\bibitem{Guth:1980zm} 
  A.~H.~Guth,
  Phys.\ Rev.\ D {\bf 23}, 347 (1981)
  
\bibitem{Linde:1981mu} 
  A.~D.~Linde,
  Phys.\ Lett.\ B {\bf 108}, 389 (1982).
  
\bibitem{Albrecht:1982wi} 
  A.~Albrecht and P.~J.~Steinhardt,
  Phys.\ Rev.\ Lett.\  {\bf 48}, 1220 (1982).
  

\bibitem{Linde:2005ht}
A.~D.~Linde,
Contemp. Concepts Phys. \textbf{5} (1990), 1-362
[arXiv:hep-th/0503203 [hep-th]].

\bibitem{Albrecht:1982mp}
A.~Albrecht, P.~J.~Steinhardt, M.~S.~Turner and F.~Wilczek,
Phys. Rev. Lett. \textbf{48} (1982), 1437

\bibitem{Abbott:1982hn}
L.~F.~Abbott, E.~Farhi and M.~B.~Wise,
Phys. Lett. B \textbf{117} (1982), 29

\bibitem{Berera:1995wh}
A.~Berera and L.~Z.~Fang,
Phys. Rev. Lett. \textbf{74} (1995), 1912-1915
[arXiv:astro-ph/9501024 [astro-ph]].

\bibitem{Berera:1996fm}
A.~Berera,
Phys. Rev. D \textbf{55} (1997), 3346-3357
[arXiv:hep-ph/9612239 [hep-ph]].

\bibitem{Berera:1999ws}
A.~Berera,
Nucl. Phys. B \textbf{585} (2000), 666-714
[arXiv:hep-ph/9904409 [hep-ph]].

\bibitem{Berera:2008ar}
A.~Berera, I.~G.~Moss and R.~O.~Ramos,
Rept. Prog. Phys. \textbf{72} (2009), 026901
[arXiv:0808.1855 [hep-ph]].

\bibitem{Bartrum:2013fia}
S.~Bartrum, M.~Bastero-Gil, A.~Berera, R.~Cerezo, R.~O.~Ramos and J.~G.~Rosa,
Phys. Lett. B \textbf{732} (2014), 116-121
[arXiv:1307.5868 [hep-ph]].

\bibitem{Dymnikova:2000gnk}
I.~Dymnikova and M.~Khlopov,
Mod. Phys. Lett. A \textbf{15} (2000), 2305-2314
[arXiv:astro-ph/0102094 [astro-ph]].

\bibitem{Panotopoulos:2015qwa}
G.~Panotopoulos and N.~Videla,
Eur. Phys. J. C \textbf{75} (2015) no.11, 525
[arXiv:1510.06981 [gr-qc]].

\bibitem{Benetti:2016jhf}
M.~Benetti and R.~O.~Ramos,
Phys. Rev. D \textbf{95} (2017) no.2, 023517
[arXiv:1610.08758 [astro-ph.CO]].

\bibitem{Motaharfar:2018mni}
M.~Motaharfar, E.~Massaeli and H.~R.~Sepangi,
JCAP \textbf{10} (2018), 002
[arXiv:1807.09548 [gr-qc]].

\bibitem{Graef:2018ulg}
L.~L.~Graef and R.~O.~Ramos,
Phys. Rev. D \textbf{98} (2018) no.2, 023531
[arXiv:1805.05985 [gr-qc]].

\bibitem{Arya:2018sgw}
R.~Arya and R.~Rangarajan,
Int. J. Mod. Phys. D \textbf{29} (2020) no.08, 2050055
[arXiv:1812.03107 [astro-ph.CO]].

\bibitem{Kamali:2018ylz}
V.~Kamali,
Eur. Phys. J. C \textbf{78} (2018) no.11, 975
[arXiv:1811.10905 [gr-qc]].

\bibitem{Samart:2021eph}
D.~Samart, P.~Ma-adlerd and P.~Channuie,
Eur. Phys. J. C \textbf{82} (2022) no.2, 122
[arXiv:2105.14552 [gr-qc]].

\bibitem{Samart:2021hgt}
D.~Samart, P.~Ma-adlerd, P.~Koad and P.~Channuie,
[arXiv:2109.09153 [astro-ph.CO]].

\bibitem{Amake:2021bee}
W.~Amake, A.~Payaka and P.~Channuie,
[arXiv:2111.07141 [gr-qc]].


\bibitem{Berera:1998gx}
A.~Berera, M.~Gleiser and R.~O.~Ramos,
Phys. Rev. D \textbf{58}, 123508 (1998)
[arXiv:hep-ph/9803394 [hep-ph]].

\bibitem{Berera:2001gs} 
A.~Berera and R.~O.~Ramos,
Phys. Rev. D \textbf{63}, 103509 (2001)
[arXiv:hep-ph/0101049 [hep-ph]].

\bibitem{Zhang:2009ge}
Y.~Zhang,
JCAP \textbf{03}, 023 (2009)
[arXiv:0903.0685 [hep-ph]].

\bibitem{Bezrukov:2008ut}
F.~Bezrukov, D.~Gorbunov and M.~Shaposhnikov,
JCAP \textbf{06} (2009), 029
[arXiv:0812.3622 [hep-ph]].

\bibitem{Bastero-Gil:2016qru}
M.~Bastero-Gil, A.~Berera, R.~O.~Ramos and J.~G.~Rosa,
Phys. Rev. Lett. \textbf{117}, no.15, 151301 (2016)
[arXiv:1604.08838 [hep-ph]].

\bibitem{Nojiri:2010wj}
S.~Nojiri and S.~D.~Odintsov,
Phys. Rept. \textbf{505} (2011), 59-144
[arXiv:1011.0544 [gr-qc]].

\bibitem{Nojiri:2017ncd}
S.~Nojiri, S.~D.~Odintsov and V.~K.~Oikonomou,
Phys. Rept. \textbf{692} (2017), 1-104
[arXiv:1705.11098 [gr-qc]].

\bibitem{Sotiriou:2008rp}
T.~P.~Sotiriou and V.~Faraoni,
Rev. Mod. Phys. \textbf{82} (2010), 451-497
[arXiv:0805.1726 [gr-qc]].

\bibitem{DeFelice:2010aj}
A.~De Felice and S.~Tsujikawa,
Living Rev. Rel. \textbf{13} (2010), 3
[arXiv:1002.4928 [gr-qc]].

\bibitem{Fujii2003} 
Y.~Fujii and K.~Maeda, ``The scalar-tensor theory of gravitation'', Cambridge University Press, 2003.

\bibitem{Maeda:1988ab}
K.~i.~Maeda,
Phys. Rev. D \textbf{39} (1989), 3159

\bibitem{Motohashi:2014tra}
H.~Motohashi,
Phys. Rev. D \textbf{91} (2015), 064016
[arXiv:1411.2972 [astro-ph.CO]].

\bibitem{Renzi:2019ewp}
F.~Renzi, M.~Shokri and A.~Melchiorri,
Phys. Dark Univ. \textbf{27} (2020), 100450
[arXiv:1909.08014 [astro-ph.CO]].

\bibitem{Liu:2018htf}
L.~H.~Liu,
[arXiv:1807.00666 [gr-qc]].

\bibitem{Rojas:2022dky}
C.~Rojas,
[arXiv:2203.00741 [gr-qc]].

\bibitem{Planck:2018jri}
Y.~Akrami \textit{et al.} [Planck],
Astron. Astrophys. \textbf{641} (2020), A10
[arXiv:1807.06211 [astro-ph.CO]].

\bibitem{Graham2009}
C.~Graham and I.~G.~Moss,
JCAP \textbf{07} (2009), 013
[arXiv:0905.3500 [astro-ph.CO]].

\bibitem{Bastero-Gil:2018uep}
M.~Bastero-Gil, A.~Berera, R.~Hern\'andez-Jim\'enez and J.~G.~Rosa,
Phys. Rev. D \textbf{98} (2018) no.8, 083502
[arXiv:1805.07186 [astro-ph.CO]].

\bibitem{BasteroGil:2009ec}
M.~Bastero-Gil and A.~Berera,
Int. J. Mod. Phys. A \textbf{24}, 2207-2240 (2009)
[arXiv:0902.0521 [hep-ph]].


\bibitem{Hall:2003zp}
L.~M.~H.~Hall, I.~G.~Moss and A.~Berera,
Phys. Rev. D \textbf{69}, 083525 (2004)
[arXiv:astro-ph/0305015 [astro-ph]].

\bibitem{Taylor:2000ze}
A.~N.~Taylor and A.~Berera,
Phys. Rev. D \textbf{62}, 083517 (2000)
[arXiv:astro-ph/0006077 [astro-ph]].

\bibitem{Moss:2008yb}
I.~G.~Moss and C.~Xiong,
JCAP \textbf{11} (2008), 023
[arXiv:0808.0261 [astro-ph]].

\bibitem{Ramos:2013nsa}
R.~O.~Ramos and L.~A.~da Silva,
JCAP \textbf{03}, 032 (2013)
[arXiv:1302.3544 [astro-ph.CO]].

\bibitem{DeOliveira:2001he}
H.~P.~De Oliveira and S.~E.~Joras,
Phys. Rev. D \textbf{64}, 063513 (2001)
[arXiv:gr-qc/0103089 [gr-qc]].

\bibitem{Visinelli:2016rhn}
L.~Visinelli,
JCAP \textbf{07}, 054 (2016)
[arXiv:1605.06449 [astro-ph.CO]].

\bibitem{Codello:2014sua}
A.~Codello, J.~Joergensen, F.~Sannino and O.~Svendsen,
JHEP \textbf{02} (2015), 050
[arXiv:1404.3558 [hep-ph]].


\bibitem{Bastero-Gil:2011rva}
M.~Bastero-Gil, A.~Berera and R.~O.~Ramos,
JCAP \textbf{07} (2011), 030
[arXiv:1106.0701 [astro-ph.CO]].

\bibitem{Bastero-Gil:2012akf}
M.~Bastero-Gil, A.~Berera, R.~O.~Ramos and J.~G.~Rosa,
JCAP \textbf{01} (2013), 016
[arXiv:1207.0445 [hep-ph]].

\bibitem{Bastero-Gil:2010dgy}
M.~Bastero-Gil, A.~Berera and R.~O.~Ramos,
JCAP \textbf{09} (2011), 033
[arXiv:1008.1929 [hep-ph]].

\bibitem{Berghaus:2019whh}
K.~V.~Berghaus, P.~W.~Graham and D.~E.~Kaplan,
JCAP \textbf{03} (2020), 034
[arXiv:1910.07525 [hep-ph]].

\bibitem{Laine:2021ego}
M.~Laine and S.~Procacci,
JCAP \textbf{06} (2021), 031
[arXiv:2102.09913 [hep-ph]].

\bibitem{Motaharfar:2021egj}
M.~Motaharfar and R.~O.~Ramos,
Phys. Rev. D \textbf{104} (2021) no.4, 043522
[arXiv:2105.01131 [hep-th]].

\bibitem{BasteroGil:2011xd}
M.~Bastero-Gil, A.~Berera and R.~O.~Ramos,
JCAP \textbf{07}, 030 (2011)
[arXiv:1106.0701 [astro-ph.CO]].

\bibitem{deOliveira:1997jt}
H.~P.~de Oliveira and R.~O.~Ramos,
Phys. Rev. D \textbf{57} (1998), 741-749
doi:10.1103/PhysRevD.57.741
[arXiv:gr-qc/9710093 [gr-qc]].

\end{thebibliography}
\end{document}